\documentclass[preprint,showpacs,amsmath,amssymb]{revtex4}

\usepackage{graphicx}
\usepackage{dcolumn}
\usepackage{bm}

\include{epsf} 

\begin{document}

\title{Tsallis nonextensive statistical mechanics of El Ni$\tilde{n}$o
Southern Oscillation Index}

\author{M. Ausloos} 
\author{F. Petroni}

\altaffiliation[Also at ]{Dipartimento di Matematica, Universit\'a
dell'Aquila, 67010 L'Aquila, Italy}
 
\affiliation{SUPRATECS, B5,
Sart Tilman, \\ B-4000 Li\`ege, Euroland}

\begin{abstract}
The shape and tails of partial distribution functions (PDF) for a climatological
signal, i.e. the El Ni$\tilde{n}$o SOI and the turbulent nature of the
ocean-atmosphere variability are linked through a model encompassing Tsallis
nonextensive statistics and leading to evolution equations of the Langevin and
Fokker-Planck type. A model originally proposed to describe the intermittent
behavior of turbulent flows describes the behavior of the normalized variability for
such a climatological index, for small and large time windows, both for small and
large variability. This normalized variability distributions can be sufficiently
well fitted with a $\chi^2$-distribution. The transition between the small time
scale model of nonextensive, intermittent process and the large scale Gaussian
extensive homogeneous fluctuation picture is found to occur at  above $ca.$ a 48~months
time lag. The intermittency exponent ($\kappa$) in the framework of the
Kolmogorov log-normal model is found to be related to the scaling exponent of the
PDF moments. The value of $\kappa$ (= 0.25) is in agreement with the intermittency exponent recently obtained for other atmospheric data.
\end{abstract}



\maketitle

\section{Introduction}

Fractional Gaussian noises and fractional Brownian motions \cite{r1} have served
recently as models for a wide variety of data in various fields, like meteorology
\cite{kimaeeta,bunde}, geology \cite{turcotte}, cardiac dynamics
\cite{plamennature1996}, finance \cite{3,4,5}. The concept of fractional Gaussian
noise (fGn) as formulated by Mandelbrot and Van Ness \cite{r1} is a derivative
process obtained from fractional Brownian motion (fBm) $B_H(t)$, namely $\lim
_{\delta \longrightarrow 1}(B_H(t+\delta)-B_H(t))/\delta$. As it has been shown
by Flandrin \cite{flandrin}. Even though the fBm is a non-stationary process it
obeys a power law over all frequencies. In this study we use fGn as a model for a
climatological signal.

The origin of non-Gaussian, thus non negligible large and sometimes so called  extreme volatility events
characterized by so called fat tailed distributions is a key question in
statistical physics; the fat tails of (short and long-range) volatilities are
thought to be caused by some `dynamical process'. Destroying all correlations,
e.g. by shuffling the order of the fluctuations, is known to cause the fat tails
almost to vanish. 

The fat tails indicate an unexpected high probability of large changes. These
extreme events are of utmost importance for risk analysis. They are considered to
be a set of strong bursts in the energy dissipation. In so doing the PDF and the
fat tail event existence are thought to be similar to the notion of intermittency
in turbulent flows \cite{neto}. 

It is an open question whether both the fat-tailed power-law of partial
distribution functions (PDF) of the various volatilities and their $evolution$
for {\it different time delays} in climatological indices can be described.

On the other hand, the non-Gaussian character of the fully developed turbulence
\cite{frisch} has been linked to nonextensive statistical physics
\cite{tsallis,tsallis_bukman,tsallis_wilk,tsallis_arim,tsallis_beck1,
tsallis_beck2,tsallis_beck3,Sattin1,Sattin2}. It seems that there is no study of
Tsallis statistics application or approach in climatology \cite{EGU}. 

One of the most intriguing phenomena in climatology, known as {\it El Ni$\tilde{n}$o}, i.e.
the more or less cyclic warming and cooling of the eastern and central regions in
the Pacific Ocean, appears to be very complicated to describe \cite{Vallis,models,ghil+98,kepp+92}.
There are three sorts of models, based on special types of filters designed
specifically to detect a signal from given atmospheric input. The ability of
these models for producing successful forecasts of El Ni$\tilde{n}$o appears to
be concomitant to the {\it very low-frequency and large-scale evolution} of the
characteristic patterns in the atmospheric boundary layer circulation. This
evolution can be thought of as a signal that precedes El Ni$\tilde{n}$o events.
It would be useful to have model-independent results with basic geophysical
inputs.

El Ni$\tilde{n}$o is a disruption of the ocean-atmospheric system in the tropical
Pacific having important consequences for weather around the globe. It is
factually described by the so called {\it Southern oscillation index (SOI)}. Much
of the drastic and tragic events occurring in North America, Tropical Africa and
Europe are also attributed to El Ni$\tilde{n}$o.  There are different ways of characterizing El Ni$\tilde{n}$o events. One of them is by the values of an index such as the Southern Oscillation Index (a proxy measure of El Ni$\tilde{n}$o based on surface air pressure differences between Darwin, Australia and Tahiti, French Polynesia) \cite{walker}, or large sea surface temperatures (SST) and sea surface height (SSH) anomalies in the eastern tropical Pacific Ocean. Here we will use the 
sea level pressure (SLP) differences between two meteorological stations, one at
Tahiti, the other at Darwin. Questions on variability of the SOI on various time
scales are relevant for better modelling. It was recently found for the southern
oscillation index (SOI), characterizing El Ni$\tilde{n}$o events
\cite{AusloosIvanova2001} that long-range correlations exist between the
fluctuations of the index. Also, correlation between the SOI and the North Atlantic Oscillation (NAO) index has been reported \cite{garcia+00} having a common oscillation of about $6-8$ years.  The NAO was studied {\it per se}, in \cite{CCMANAO}.

In this paper the behavior of a climatological index, i.e. the SOI, on $short$
and $large$ time windows (time scales to be better defined below) is studied
along the lines of a recently suggested model of hydrodynamic turbulence that
serves as a dynamic foundation for nonextensive statistics
\cite{tsallis_beck1,tsallis_beck2,tsallis_beck3}.

In Sect. 2, we describe the distribution of variability for the monthly value
signal of the SOI index for the time interval between Jan. 1866 and Jan. 2006,
thus a series of $N=$1681 months or data points, values downloaded from the
Climate Prediction Center web site ($http://www.cpc.ncep.noaa.gov$
$/data/indices/$). We also characterize the tail(s) of the distribution for various
time lags $\Delta t$'s, i.e. from 1 month up to 36~months (or 3 years!), and will
observe the value of the PDF tails, for such time lags, outside the best gaussian
fit through the data. There are several ways of displaying features in the
variability of a climatological index. A simple one represents value increment
$\Delta y(t)=y(t+\Delta t) - y(t)$ or difference between the value of the index
$y(t)$ at time $t+\Delta t$ and its value at time $t$. Below we mainly consider
the {\it normalized variability} $Z(t,\Delta t)=(\Delta y(t) - <\Delta y>_{\Delta
t})/\sigma_{\Delta t}$, where $<\Delta y>_{\Delta t}$ denotes the average and
$\sigma_{\Delta t}$ the standard deviation of $\Delta y(t)$ for a given $\Delta
t$. The normalized variability $Z(t,\Delta t)$ depend on the time $t$ and the
time lag $\Delta t$. However, in order to simplify the notations and whenever
possible without leading to confusion and misunderstanding we will drop the
explicit writing of one or both variables.

In Sect. 3, we calculate the power law exponents characterizing the $integrated$
distribution of the normalized variability over different time lags for the SOI
index monthly values through a detrended fluctuation analysis and a power
spectral density analysis point of view.

In all cases it is useful to test the null hypothesis or estimate the error bars
with respect to standard signals. It is thought \cite{neto} that the fat tails
are caused by long-range volatility correlations. Destroying all correlations by
shuffling the order of the fluctuations, is known to cause the fat tails almost
to vanish. A Kolmogorov-Smirnov test (not shown) on shuffled data has indicated
us the statistical validity of the numerical values and the statistically
acceptable meaning of the displayed error bars. Results are compared to shuffled
data for estimating the value of the error bars.

In Sect. 4, Tsallis statistical approach is outlined, and distributions of
(normalized) variability for time lags between $\Delta t$ = 1 to 36~months are
examined. The nonextensivity, i.e. some anomalous scaling of classically
extensive properties like the entropy, is linked to a single parameter $q$, e.g.
in the Tsallis formulation of nonextensive thermostatistics. It is found that the
$q$-value of the nonextensive entropy converges to a value = 1.01 for $\Delta t$
= 36~months, starting with $q$ = 1.25 for $\Delta t$ = 1. The probability density
$f_{\Delta t}(\beta)$ of the volatility $\beta$ in terms of the standard
deviation of the normalized variability of the SOI for different time lags is
found to obey the $\chi^2$-distribution. The intermittency exponent ($\kappa$) of
the Kolmogorov log-normal model is found to be related to the scaling exponent of
the PDF moments, -thereby giving weight to this model.

In Sect. 5, the usual Fokker-Planck approach for treating the time-dependent
probability distribution functions is summarized. Coefficients governing both the
Fokker-Planck equation for the distribution function of normalized variability
and the Langevin equation for the time evolution of normalized variability of
monthly value signal of SOI are obtained. Therefore we present for the first time
a coherent theory linking the shape and tails of partial distribution functions
for long and short time lags of the monthly values of a climatological signal and
connect the often suggested turbulent nature of the ocean-atmospheric interface
interactions to a model encompassing nonextensive statistics and evolution
equations of the Langevin and Fokker-Planck type.

We will often compare results based on normalized variability and non normalized
variability time series.

\section{Data and distribution of variability} 

Monthly values of the SOI index for the time interval between Jan. 1866 and Jan.
2006 were downloaded from the Joint Institute for Study of the Atmosphere and
Ocean (JISAO) web site $http://tao.atmos.washington.edu/$
$pacs/additional_{-}analyses/soi.html$ for the longest period available there,
i.e. from Jan. 1866 to June 1999. Data for the time interval from July 1999 to
Jan. 2006 were downloaded from the Climate Prediction Center NCEP web site
($http://www.cpc.ncep.noaa.gov/data/indices/$). 
For the years before 1866 daily measurements of the sea level pressure at both
stations have been reported to exist and monthly values of the southern
oscillation index have been calculated \cite{jones} back to 1841. However, there
are gaps of a couple of years in the record whence not suitable for our analysis. 
We have chosen the JISAO's data for the period before July 1999, since they do
not contain missing values as the data series from NCEP do. The data are fully compatible
because the standardization of both JISAO and NCEP data series is calculated through
the standard deviation $S$ of the sea level pressure (SLP) at a station in Tahiti
and the sea level pressure at a station in Darwin  

\begin{equation} SOI = \frac{P_{Tahiti} - P_{Darwin}}{S}. \end{equation}

Such SOI monthly data are
plotted as a function of time in Fig. 1. The data set consists of $1681$ data points. It is sometimes stated that daily or weekly values of the SOI do not convey much
in the way of useful information about the current state of the climate, and
accordingly the Bureau of Meteorology does not issue them. Daily values in
particular can fluctuate markedly because of daily weather patterns, and should
not be used for climate purposes. We may disagree with this statement (provided the reliability of the data). There are
indeed techniques which can sort out noise from coherent behavior
\cite{Davisbook}. 

Sustained negative values of the SOI often indicate El Ni$\tilde{n}$o episodes \cite{bomgovau}.
These negative values are usually accompanied by sustained warming of the central
and eastern tropical Pacific Ocean, a decrease in the strength of the Pacific
Trade Winds, and a reduction in rainfall over eastern and northern Australia. The
most recent strong El Ni$\tilde{n}$o was in 1997/98. Positive values of the SOI
are associated with stronger Pacific trade winds and warmer sea temperatures to
the north of Australia, popularly known as a La Ni$\tilde{n}$a episode. Waters in
the central and eastern tropical Pacific Ocean become cooler during this time.
Together these give an increased probability that eastern and northern Australia
will be wetter than normal. The most recent strong La Ni$\tilde{n}$a was in
1988/89; a moderate La Ni$\tilde{n}$a event occurred in 1998/99, which weakened
back to neutral conditions before reforming for a shorter period in 1999/2000.
This last event finished in Autumn 2000.

The distribution of the normalized variability $Z(t,\Delta t)$ of the monthly
value signal of SOI index between Jan. 1866 and Jan. 2006, for $\Delta t$ =
1~month are plotted in Fig. 2a. The partial distribution of non normalized
variability $\Delta y$ of the monthly value signal of SOI index is plotted in Fig
2b as we will later compare some of our findings for both $Z$ and $\Delta y$. For
comparison A fit is first attempted with a Gaussian distribution for $small$
values of the increments, i.e. the central part of the distribution. The
distribution is well fitted with such a Gaussian type curve within the interval
$Z\in]-2,2[$ but departs from the Gaussian form outside this interval. The
negative and positive tails of the distribution outside the Gaussian curve are
found both to be equal to -5.6. In the case $\Delta t\ge$ 1~month, it is observed
that the best Gaussian range is increasing with increasing time lag (Fig. 3).

\section{Time correlations and spectral power}

There are different estimators for the long and/or short range dependence of
fluctuations correlations \cite{taqqu}.

Through the (linearly) detrended fluctuation analysis (DFA) method, see e.g.
\cite{hu}, we show first that the {\it long range correlations} of monthly value
signal of SOI for the time interval of interest, are $1/f$-like. The method has
been used previously to identify whether long range correlations exist in
non-stationary signals, in many research fields such as e.g. finance \cite{4,5},
cardiac dynamics \cite{plamennature1996} and of course meteorology
\cite{kimaeeta,bunde,buda_dfa}. Its concepts are therefore not repeated here. For
an extensive list of references see \cite{hu}. Briefly, the signal time series
$y(t)$ is $first$ $integrated$, to `mimic' a random walk $Y(t)$. The time axis (form 1 to $N$)
is next divided into non-overlapping boxes of equal size $n$; one
looks thereafter for the best (linear) trend, $z_n$, in each box, and calculates
the root mean square deviation of the (integrated) signal with respect to $z_n$
in each box. The average of such values is taken at fixed box size $n$ in order
to obtain

\begin{equation} F(n) = \sqrt{{1 \over N } {\sum_{i=1}^{N}
{\left[Y(i)-z_n(i)\right]}^2}} \end{equation}

\noindent The box size is next varied over the $n$ value. The resulting function
is expected to behave like $ n^{1+H_{DFA}}$ indicating a scaling law. For the
(integrated) monthly value signal of the SOI index, a scaling exponent
$1+H_{DFA}=1.05\pm0.01$ is found (Fig. 4) in a scaling range extending from about
4 to about 66-72~months. A signal with Hausdorff dimension $H_{DFA}$ close to zero
has the characteristics of a fractional Gaussian noise signal
\cite{turcotte,percival}. 

Along the same line of thought the scaling properties of the normalized
variability $Z(t,\Delta t)=(\Delta y(t) - <\Delta y>_{\Delta t})/ \sigma_{\Delta
t}$ have also been tested for different time lag values, i.e. $\Delta
t=1,3,6,12,24,36$~months (Fig. 5). The DFA functions, as defined here above, of
the integrated normalized variability shows non trivial scaling properties  for
the series of normalized variability. The
values of the scaling exponents and the maximum box size $n_x$ (in days) for
which the scaling holds for each DFA-function are given in Table \ref{table1},
while the DFA-functions together with fitting lines are plotted in Fig. 5.

The power spectrum of the monthly value signal of SOI $S(f)\sim f^{-\mu}$ with
spectral exponents $\mu_1=-0.26$ and $\mu_2=1.20$ with a scale break at
1/70~months$^{-1}$ is shown in Fig. 6. The scaling properties of the power
spectrum of two surrogate data, one in which the amplitudes are randomly shuffled
and another in which the magnitudes are preserved but the sign of the data is
shuffled, are shown in the inserts of Fig. 6. Such scaling spectral exponents
$\mu=0$ are signature of a white noise like behavior. Recall that $\mu =2.0$
corresponds to usual Brownian motion. The theoretical relationship $\mu=2 H_{DFA}-1$
is approximately verified, - the weak agreement being likely due to the quite limited data size.

We have also checked for scaling behavior and possible periodicities in the power
spectrum of the time series of the normalized variability $Z(t,\Delta t)=(\Delta
y(t) - <\Delta y>_{\Delta t})/\sigma_{\Delta t}$ for different (selected) values
of the time lag $\Delta t=1,3,6,12,24,36$~months (Fig. 7).

Periodicities in the power spectrum of the normalized variability time series for
$\Delta t>1$~month were expected to be found since these periods are somewhat
embedded into the time series by the way they are obtained and the Fourier
transform technique. It is easily observed that the maxima and the minima of the
spectrum correspond to harmonics and subharmonics of $1/\Delta t$.

\section{Tsallis statistics}

Based on the scaling properties of multifractals \cite{mf}
Tsallis \cite{tsallis,tsallis_1995} proposed a generalized Boltzmann-Gibbs
thermo-statistics through the introduction of a family of non-extensive entropy
functional ${\cal S}_q$ given by:

\begin{equation} {\cal S}_q=k \frac{1}{q-1}\left(1-\int p(x,t)^q dx\right),
\label{ts} \end{equation} with a single parameter $q$ and where $k$ is a
normalization constant. The main ingredient in Eq.(\ref{ts}) is the
time-dependent probability distribution $p(x,t)$ of the stochastic variable $x$.
The functional is reduced to the classical extensive Boltzmann-Gibbs form in the
limit of $q\longrightarrow 1$. The Tsallis parameter $q$ characterizes the
non-extensivity of the entropy. Subject to certain constraints the functional in
Eq.(\ref{ts}) seems to yield a probability distribution function of the form
\cite{neto,tsallis,tsallis_beck1,michael,kozuki}

\begin{equation} p(x) = \frac{1}{Z_q}\left\{1+\frac{C \beta_0
2\alpha(q-1)|x|^{2\alpha}} {2\alpha-(q-1)}\right\}^{-\frac{1}{(q-1)}}
\label{tsallis} \end{equation} for the stochastic variable $x$, where

\begin{equation} \frac{1}{Z_q}=\alpha\left\{\frac{C \beta_0 2\alpha(q-1)}
{2\alpha-(q-1)}\right\}^{1/2\alpha} \,
\frac{\Gamma\left(\frac{1}{q-1}\right)}{\Gamma\left(\frac{1}{2\alpha}\right)
\Gamma\left(\frac{1}{q-1}-\frac{1}{2\alpha}\right)} \label{zq} \end{equation} in
which $C$ is a constant and $0<\alpha\le 1$ is the power law exponent of the
potential $U(x)=C|x|^{2\alpha}$ that provides the `restoring force' $F(x)$ in
Beck model of turbulence
\cite{tsallis_beck1,tsallis_beck2,tsallis_beck3,Sattin2}. The latter is described
by a Langevin equation

\begin{equation} \frac{dx}{dt}=-\gamma F(x)+R(t) \label{force} \end{equation}
where $\gamma$ is a parameter and $R(t)$ is a gaussian white noise. A non-zero
value of $\gamma$ corresponds to providing energy to (or draining from) the
system by the outside \cite{Sattingrangas}. The parameter $\beta_0$ in
Eq.(\ref{tsallis}) and (\ref{zq}) is the mean of the fluctuating standard
deviation $\beta$, i.e. the local standard deviation of $|x|$ over a certain
window of size $m$ \cite{neto}. We will use this model assuming that the
normalized variability $Z(t,\Delta t)$ represent $the$ stochastic variable $x$,
as in Eq.(1). We will search whether Eq.(\ref{tsallis}) is obeyed for $x\equiv
Z(t,\Delta t)$, thus studying $p(x)\equiv p_{\Delta t}(Z)$ for various time lags
$\Delta t$.

Just as in Beck model of turbulence
\cite{tsallis_beck1,tsallis_beck2,tsallis_beck3} we assume that the standard
deviation $\beta$ is $\chi^2$-distributed with degree $\nu$ (see another formula
in \cite{Sattin2}):

\begin{equation} f_{\Delta t}(\beta) \equiv
\frac{1}{\Gamma(\nu/2)}\left(\frac{\nu}{2\beta_0} \right)^{\nu/2}
\beta^{\nu/2-1}\exp\left(-\frac{\nu\beta}{2\beta_0}\right), \quad \nu > 2,
\label{chi} \end{equation} where $\Gamma$ is the Gamma function,
$\beta_0=<\beta>$ and the number of degrees of freedom $\nu$ can be found from:
\begin{equation} \nu=\frac{2<\beta>^2}{<\beta^2>-<\beta>^2}. \label{nu}
\end{equation}

The Tsallis parameter $q$ satisfies \cite{tsallis_beck1} \begin{equation} q\equiv
1 + \frac{2\alpha}{\alpha \nu +1}. \label{q} \end{equation}

To justify our assumption that the `local' standard deviation of the normalized
variability $Z(t,\Delta t)$ is of the form of $\chi^2$-distribution, we checked
the distribution of the normalized variability of the monthly value signal of
SOI. We have calculated the standard deviation of the normalized variability
within various non-overlapping windows of size $m$, ranging from 6 to 36~months

\begin{equation} \beta(k)=\sqrt{{1 \over m}\sum_{i=km+1}^{(k+1)m}Z^2(i) -
\left({1 \over m}\sum_{i=km+1}^{(k+1)m}Z(i)\right)^2} \label{volatility}
\end{equation}

In doing so we have a various number of ${\cal M}$ non-overlapping windows for
various time lags $\Delta t$, and have searched for the most efficient size of
the window in order not to loose data points and therefore, information. The
resulting empirically obtained distributions of the `local' standard deviation
(Eq.(\ref{volatility})) of normalized variability for the different time lags of
interest are plotted in Fig. 8 for an intermediary case $m=12$. The values of the
degree $\nu$ of the $\chi^2$-distribution are then obtained using Eq.~(\ref{nu}).
The spread $[\beta_{min},\beta_{max}]$ of the local standard deviation $\beta$
decreases with increasing the time lag as it is expected from a
$\chi^2$-distribution function due to the exponential function in Eq. (\ref{chi})
for large values of the degree of freedom $\nu$. The value of $\nu$ much varies
as a function of $m$ and the time lags considered. The fits are always very good.
However the $\beta_0$ and $\nu$ values are quite dependent on the parameters used
in the numerical analysis. Based on these results, e.g. Fig. 8, it can be
accepted that the (turbulent) model $\beta$-distributions can be sufficiently
well fitted for our purpose with a $\chi^2$-distribution, thereby justifying the
initial assumption.\footnote{Sattin formula \cite{Sattin2} might also be tested
in future work.}

The probability distributions of the normalized variability for the different
values of the time lag $\Delta t=1,3,6,12,24,36$~months are shown in Fig. 3
together with the lines representing the best fit to the Tsallis type of
distribution function. In Table \ref{table2} the statistical parameters related
to the Tsallis type of distribution function are summarized, including a
criterion for the goodness of the fit, i.e. the Kolmogorov-Smirnov distance
$d_{KS}$, which is defined as the maximum distance between the cumulative
probability distributions of the data and the fitting lines. Note that the
kurtosis (see Table \ref{table2}) for the Tsallis type of distribution function
\begin{equation} K_r=K_L\frac{(5-3q)}{(7-5q)}, \label{kr} \end{equation} where
$K_L=3$ for a Gaussian process, is positive for all values of $q<7/5$ as
expected, since its positiveness is directly related to the occurrence of
intermittency \cite{neto}. Moreover, the limit $q<7/5$ also implies that the
second moment of the Tsallis type distribution function will always remain
finite, as necessarily due to the type of phenomena hereby studied.

Furthermore, if we assume that the Kolmogorov log-normal model of turbulence
\cite{k62} is applicable and let $\Delta t_L$ be the scale at which the $whole$
partial distribution function becomes Gaussian, then the kurtosis $K_r$ should
scale as

\begin{equation} K_r=K_L\left(\frac{\Delta t}{\Delta t_L}\right)^{-\delta}
\label{krL}. \end{equation} Therefore 

\begin{equation} q=\frac{5 - 7 \left(\Delta t/\Delta t_L\right)^{-\delta}} {3 - 5
\left(\Delta t/\Delta t_L\right)^{-\delta}}. \label{qkr} \end{equation} 

In order to obtain an estimate for $\Delta t_L$, we increase the time lag to the
value $\Delta t=48$~months, quite outside the range so far examined (see Fig.3
for example) leading to a rather complete coincidence between the distribution
functions in the Tsallis and Gaussian forms for the presently investigated data.
The corresponding parameter values are listed in Table \ref{table2}. A quick
perusal observation convincingly indicates where the transition occurs between
the small time scale model of nonextensive, intermittent process and the large
scale Gaussian extensive homogeneous fluctuation picture \cite{neto,tsallis}.

In Fig. 9 the Tsallis parameter $q$ is shown as a function of the rescaled time
lags $\Delta t/\Delta t_L$, where $\Delta t_L$ is the integral scale, the scale
at which the $whole$ probability distribution function converges to Gaussian. The
crosses represent the $q$ values for which the best fit to the SOI data (Fig. 3)
is obtained with Eq. (\ref{tsallis}). With this the value of the integral scale
$\Delta t_L$, we find the value of the exponent $\delta= 0.11$ as the one for
which the Eq. (\ref{qkr}) fits best the $q$-values. The exponent value $\delta=
0.11$ also allows to fit well the power law dependence (Eqs. (\ref{kr}) and
(\ref{krL})) of the rescaled kurtosis $K_r/K_L$ as shown in the insert of Fig. 9.

Note that in the framework of the Kolmogorov log-normal model
\cite{k62,tsallis_beck2}, $\delta= 4\kappa/9$, where $\kappa$ is called the
intermittency exponent. Therefore, we find $\kappa=0.25$ for the intermittency
exponent of normalized variability of the SOI signal in the time interval of
interest. This value of $\kappa$ is $interestingly$ the same as the value of the intermittency
exponent $\kappa=0.25$ for turbulence recently obtained from experimental
atmospheric data \cite{sree_update}. Early estimates have varied from 0.18 to
0.85 using different experimental techniques \cite{sree_temp,sree_mf,wyngaard}.
Large range of values of the intermittency exponent, ranging from 0.2 to 0.8,
have been reported in studies of multiparticle production \cite{janik}. It was
found that the range of intermittency exponent values depend on the number of
cascades; the smaller the number of stages of the multiplicative cascade the
smaller $\kappa$, and conversely [Fig. 2b in \cite{janik}].

One can explore the Tsallis type of the probability distribution function
Eq.(\ref{tsallis}) in two limits. For small values of normalized log variability
$Z$ the probability distribution function converges to the form

\begin{equation} p_{\Delta t}(Z) \approx \frac{1}{Z_q}\exp\left\{ -\frac{C
\beta_02\alpha}{2\alpha-(q-1)}|Z|^{2\alpha}\right\} \label{ts_gauss}
\end{equation} Therefore the Tsallis type distribution function converges to a
Gaussian, i.e. $\alpha \longrightarrow 1$, for small values of the normalized log
variability, for any $\Delta t$ investigated hereby (see Figs. 2-3).

In the limit of large values of normalized variability $Z$, the Tsallis type
distribution converges to a power law

\begin{equation} p_{\Delta t}(Z) \approx \frac{1}{Z_q}\left\{\frac{(q-1)C \beta_0
2\alpha}{2\alpha-(q-1)}|Z|^{2\alpha}\right\}^{-\frac{1}{q-1}.} \label{ts_power}
\end{equation}

Studying the Tsallis type of distribution function one can obtain from
Eq.(\ref{tsallis}) an expression for the width of the Tsallis type of probability
distribution function, $2\sigma_w^2=(2\alpha-(q-1)) / (2\alpha C\beta_0(q-1))$.
In the limit of $\alpha \longrightarrow 1$ the width of the Tsallis type
distribution
$2\sigma_w^2=(3-q)/2C\beta_0(q-1)$, i.e. $\sim 2/(C\beta_0)$. It is obvious that
for large time lags $2\sigma^2_w$ tends to diverge \cite{michael}, like $\simeq
(\Delta t)^{2/(3-q)}$; this can be easily verified on a log-log plot (not shown).

In limit of $q\longrightarrow 1$ the Tsallis type distribution function converges
to Gaussian. The values of the parameters $q$, $\alpha$, $C\beta_0$, that best
fit the data using Eq.(\ref{tsallis}), and $2\sigma_w^2$ are plotted as a
function of the time lag in Fig. 10.

\section{Fokker-Planck approach}

On the other hand, the evolution of a time dependent probability distribution
function is usually described within the Fokker-Planck approach. This method
provides some further information on the correlations present in the time series
and it begins with the joint PDF's, that depend on $\cal{N}$ variables, i.e.
$p^{\cal{N}} (Z_1,\Delta t_1;...;Z_{\cal{N}},\Delta t_{\cal{N}})$. We started to
address this issue by determining the joint PDF for ${\cal{N}}=2$, i.e.
$p(Z_2,\Delta t_2; \Delta x_1, \Delta t_1)$. The symmetrically tilted character
of the joint PDF contour levels (Fig. 11) around an inertia axis with slope +1
points out to some statistical dependence, i.e. a correlation, between the
normalized variability $Z(t,\Delta t)$ of the monthly value signal of SOI.

The conditional probability function is

\begin{equation} p ( Z_{i+1},\Delta t_{i+1}|Z_{i},\Delta t_{i}) =
\frac{p(Z_{i+1},\Delta t_{i+1};Z_{i},\Delta t_{i})}{p(Z_{i}, \Delta t_{i})}
\end{equation} for $i = 1,...,{\cal{N}}-1$. For any $\Delta t_{2}$ $<$ $\Delta
t_{i}$ $<$ $\Delta t_{1}$, the Chapman-Kolmogorov equation is a necessary
condition of a Markov process, one without memory but governed by probabilistic
conditions

\begin{equation} p(Z_{2},\Delta t_{2}|Z_{1},\Delta t_{1})= \int
d(Z_{i})p(Z_{2},\Delta t_{2}|Z_{i},\Delta t_{i})p(\Delta x_{i},\Delta
t_{i}|Z_{1},\Delta t_{1}). \end{equation}

The Chapman-Kolmogorov equation when formulated in $differential$ form yields a
master equation, which can take the form of a Fokker-P1anck equation
\cite{ernst}. Let $\tau=log_2(48/\Delta t)$,

\begin{equation} \frac{d}{d\tau}p(Z,\tau )=\left[-\frac{\partial }{\partial Z}
D^{(1)}(Z,\tau )+\frac{\partial^2 }{\partial Z^{2}} D^{(2)}(Z,\tau
)\right]p(Z,\tau ) \label{fpe} \end{equation} in terms of a drift
$D^{(1)}$($Z$,$\tau $) and a diffusion coefficient $D^{(2)}$($Z$,$\tau $) (thus
values of $\tau $ represent $ \Delta t_{i}$, $i=1,...$).

The coefficient functional dependence can be estimated directly from the moments
$M^{(k)}$ (known as Kramers-Moyal coefficients) of the conditional probability
distributions:

\begin{equation} M^{(k)}=\frac{1}{\Delta \tau }\int dZ^{^{\prime }} (Z^{^{\prime
}}-Z)^{k}p(Z^{^{\prime }},\tau +\Delta \tau |Z,\tau ) \end{equation}

\begin{equation} D^{(k)}(Z,\tau )=\frac{1}{k!}\mbox{lim} M^{(k)} \end{equation}
for $\Delta \tau \rightarrow 0$. The drift coefficient $D^{(1)}$ and the
diffusion coefficient $D^{(2)}$ are well represented (Fig. 12a,b) by a line and
parabola, respectively

\begin{equation} D^{(1)} = -0.37 Z - 0.01 \label{d1n}\end{equation}

\begin{equation} D^{(2)} = 0.10 Z^2 - 0.10 Z + 0.33 \label{d2n}\end{equation} for
the normalized variability (plotted with dots).

We have compared the above values of the drift and the diffusion coefficients for
those of the drift and the diffusion coefficients for non normalized variability
(plotted with open circles) and have obtained

\begin{equation} D^{(1)} = -0.52 \Delta y - 0.02 \label{d1nn}\end{equation}

\begin{equation} D^{(2)} = 0.20 \Delta y^2 - 0.10 \Delta y + 0.24
\label{d2nn}\end{equation}

Note that the first term on the right hand side of Eq. (\ref{fpe}) is identified
\cite{carmichael} as the term generating drift behavior in the evolution of the
PDF, while the second term is responsible for the diffusion, or fluctuation term
in the PDF evolution. In the asymptotic case when the linear term in $D^{(1)}$,
i.e. the coefficient $D_1^{(1)}$ is dominating the dependence and the independent
$D_{0}^{(2)}$ term is somewhat dominating in $D^{(2)}$, then the Fokker-Planck
equation is linear, otherwise the drift and diffusion terms are intervened.
Comparing the values of the linear and independent terms in Eqs. (\ref{d1n}) and
(\ref{d1nn}), as well as the values of the quadratic, linear and independent
terms in Eqs. (\ref{d2n}) and (\ref{d2nn}) one may argue that the linear
approximation for the Fokker-Planck equation holds more convincing for the PDF
evolution of the normalized variability (Eqs. (\ref{d1n},\ref{d1nn})) as opposite
to the PDF evolution of the non normalized variability $\Delta y$.

On the other hand, it may be worthwhile to recall that the observed quadratic
dependence of the diffusion term $D^{(2)}$ is essential for the logarithmic
scaling of the intermittency parameter in studies on turbulence.

Finally, the Fokker-Planck equation for the distribution function is known to be
equivalent to a Langevin equation for the variable, i.e. $Z$ here, (within the
Ito interpretation
\cite{risken,ernst,reich,hanggi,gardiner})

\begin{equation} \frac {d}{d\tau} Z(\tau) = D^{(1)}(Z(\tau),\tau) + \eta(\tau)
\sqrt {{D^{(2)} (Z(\tau),\tau)}}, \label{lee} \end{equation} \noindent where
$\eta(\tau)$ is a fluctuating $\delta$-correlated force with Gaussian statistics,
i.e. $<$ $\eta(\tau)$ $\eta(\tau')$$>$ = 2$ \delta (\tau -\tau')$.  

Thus the Fokker-Planck approach provides the evolution process of PDF's {\it from
small time lags to larger ones}. The fact that the drift coefficient is finite
implies that there is some `restoring force', i.e. $\gamma\ne 0$ in Eq.
(\ref{force}), while the quadratic dependence of $D^{(2)}$ in $Z$ is obviously
like an autocorrelation function for a diffusion process.

An interaction that can produce such a `restoring force' is the the air-sea
interaction that takes place in many different ways. However, its main
ingredients are the air-sea fluxes of mass, heat and momentum. Reliable estimates
of the air-sea fluxes of heat and momentum are vital to improve our understanding
of the coupled ocean-atmosphere system. An air-sea heat and momentum climatology,
as the one recently reported \cite{soc} can be hopefully used in future dynamic
models to relate the findings of this study to classical meteorological qualities
using observations.

\section{Conclusion}

In summary, we have presented a method that provides the evolution process of
probability distribution functions (over 140 years) for one
climatological index, i.e. the SOI. It can be first  recalled that crossovers in the DFA results  and power spectral density point to specific time scales, - in fact related to famous phenomena, like sunspots. 

We have mainly studied the evolution process of the
tails that are outside the central (Gaussian) regime,- which of course occurs only  at small variability,
thereby facilitating the understanding of the evolution of these distribution
functions in a Fokker-Planck framework.  The Gaussian regime range has been found to imply that signal correlations extend up to $ca.$ 48 months, - an interesting time lag to be considered in microscopic evolution model(s) of  El Ni$\tilde{n}$o.

It has been found that Beck turbulence model can be well applied
to describe the distributions of the standard deviation of the SOI signal normalized
variability assuming a $\chi^2$-distribution for these. In some sense this application (or generalization of the model) is justified by the fact that the ideas behind the turbulence model, based on temperature fluctuations, can be expected to be carried over  to the case in which pressure fluctuations occur in the system.

An open question in nonextensive thermostatistics studies is often raised about
the meaning, value and behavior of the non extensive exponent, or Tsallis
parameter $q$. The intermittency exponent is interestingly found to be related to the scaling
exponent of the PDF moments in the framework of Kolmogorov log-normal model,
thereby giving weight to the model and the statistical approach. 

We have also presented the turbulence-like dynamics through the Fokker-Planck and
the Langevin equations. We have (as it has been expected) found that, in the
treated case, there is some `restoring force', i.e. ($\gamma\ne 0$ in the
Langevin equation). A comparison is made between normalized variability and non
normalized variability.

Whence we have related a climatological signal behavior to Tsallis non extensive
thermodynamics approach, i.e. more precisely to a turbulence-like process, - as
climatological ocean-atmospheric interface interactions and indices were often
claimed to be seen. No need to say that this empirical modeling only $describes$ the evolution of the signal but does not $explain$ it, as a general circulation model \cite{GCM} should do. Nevertheless the time scales which are hereby observed might shine some light on  approximation validity or  the need to restrict extrapolations to realistic ranges, - including memory effects. Finally, it seems that we have thoroughly answered the often
raised question `why to look at the tails of a probability distribution function?
and what does that lead to?'.

\vspace*{2cm}

{\bf Acknowledgements}

Part of FP work 
has been supported by European Commission Project 
E2C2 FP6-2003-NEST-Path-012975  Extreme Events: Causes and Consequences.
 Part of this work results from  financing through the ARC 02-07/293 Project of the
 ULg and the COST P10 ''Physics of Risk" program which MA also thoroughly acknowledges. Critical and encouraging comments by  M. Ghil, H. Herrmann,  K. Ivanova, J. Peinke, C. Tsallis, and J. Vannimenus have been as always very valuable for improving this report.

\vskip 1cm \newpage \parindent=0pt \newpage \parindent=0pt

{\large \bf Figure Captions}

\vskip 0.5cm{\bf Figure 1} -- Monthly values of the Southern Oscillation Index
(SOI) as defined in the text reported from Jan. 1866 to Jan. 2006. Data are
downloaded from
$http://tao.atmos.washington.edu/pacs/additional_{}analyses/soi.html$ and from
$ftp://ftpprd.ncep.noaa.gov/pub/cpc/wd52dg/data/indices/soi$ after June 1999.
Data series consists of 1681 data points

\vskip 0.5cm{\bf Figure 2} -- (a) Probability distribution function of normalized
variability $Z(t,\Delta t)$ of monthly values signal of the Southern Oscillation
Index from Jan. 1866 to Jan. 2006 for $\Delta t=1$~month (symbols). $Z(t,\Delta
t)$ is defined as $Z(t,\Delta t)=(\Delta y(t) - <\Delta y>_{\Delta
t})/\sigma_{\Delta t}$, where $\Delta y(t)=y(t+\Delta t)-y(t)$ and
$\sigma_{\Delta t}$ is the standard deviation of $\Delta y(t)$ for time lag
$\Delta t$. The dashed line represents a Gaussian distribution. Inset: Power law
fit (solid line) of the negative and positive slope (-5.6 for both) of the
distribution outside the Gaussian regime, i.e. $]-2,+2[$. (b) same as (a) but for
non normalized variability $\Delta y$

\vskip 0.5cm{\bf Figure 3} -- Probability distribution function (PDF) $p_{\Delta
t}(Z)$ of normalized variability of monthly values signal of the SOI (symbols)
and the Tsallis type distribution function (lines) for different values of
$\Delta t=1,3,6,12,24,36$~months. The PDF (symbols and curves) for each $\Delta
t$ are moved down by 10 with respect to the previous one; the curve for $\Delta
t=1$~month is unmoved. The large dots mark the ends of the interval in which the
distribution is like a gaussian distribution. The values of the parameters for
the Tsallis type distribution function for each $\Delta t$ are summarized in
Table \ref{table2}

\vskip 0.5cm{\bf Figure 4} -- DFA function $F(n)$ plotted as a function of the
the box size $n$ of the integrated monthly values signal of the SOI from Jan.
1866 to Jan. 2006. $1/f$-like fluctuations with $slope=1.06\pm0.01$ are
obtained for time scales below 66 months and fractional Gaussian noise like
fluctuations $slope=0.36\pm0.02$ above 72 months. Insets: White noise like
fluctuations of two types of surrogate data, when the data are shuffled randomly
and when the sign of the data is shuffled randomly

\vskip 0.5cm{\bf Figure 5} -- DFA function $F(n)$ plotted as a function of the
box size $n$ of the integrated normalized variability $Z(t,\Delta t)$ of the
monthly values signal of the SOI from Jan. 1866 to Jan. 2006, for different time
lags $\Delta t=1,3,6,12,24,36$~months. Values of the scaling exponents $H_{DFA}$
for the various DFA functions are summarized in Table \ref{table1}

\vskip 0.5cm{\bf Figure 6} -- Power spectrum $S(f)$ of the monthly values signal
of the SOI from Jan. 1866 to Jan. 2006. A scale break at around
$f=1/70$~month$^{-1}$ separates two scaling regions. Insets: Scaling of the power
spectrum of both shuffled amplitude and shuffled sign of monthly values signal of the SOI
as a white noise signal with $\mu\approx 0$

\vskip 0.5cm{\bf Figure 7} -- Power spectrum $S(f)$ of the normalized variability
$Z(t,\Delta t)$ of the monthly values signal of the SOI from Jan. 1866 to Apr.
2003 for different time lags $\Delta t=1,3,6,12,24,36$~months. Each curve is
moved down by $10^{-5}$ with respect to the previous one; the power spectrum of
the normalized returns for $\Delta t=1$~month is not displaced

\vskip 0.5cm{\bf Figure 8} -- Probability density $f_{\Delta t}(\beta)$ of the
local volatility $\beta$ (Eq.(\ref{volatility})) in terms of standard deviation
of the normalized variability $Z(t,\Delta t)$ of SOI in non-overlapping windows
with size $m$=12~months for different time lags (symbols) (a-f) $\Delta
t=1,3,6,12,24,36$~months. Lines: $\chi^2$-distribution as given by Eq.
(\ref{chi})

\vskip 0.5cm{\bf Figure 9} -- The functional dependence of the Tsallis $q$
parameter on the rescaled time lag $\Delta t/\Delta t_L$ for $\Delta
t_L=48$~months and $\delta= 0.11$ (see Eq. (\ref{qkr})) (line); the symbols
represent the values of the $q$ parameter listed in Table \ref{table2} and used
to plot the fitting lines in Fig. 2. Inset : Scaling properties of the rescaled
kurtosis $K_r/K_L$, where $K_L=3$ is the kurtosis for a Gaussian process, as a
function of the rescaled time lag $\Delta t/\Delta t_L$ satisfying Eq. (\ref{kr})
(open symbols) and Eq. (\ref{krL}) (full symbols)

\vskip 0.5cm{\bf Figure 10} -- Characteristic parameters of Tsallis type
distribution function as defined in \cite{kozuki} : Tsallis $q$-parameter
(crosses), $\alpha$ (squares), constant $C\beta_0$ used in the fit (open
circles), the width of the Tsallis type distribution $2\sigma_w^2=(2\alpha-(q-1))
/ (2\alpha C\beta_0(q-1))$ from Eq.(\ref{tsallis}) (triangles) (rescaled by a
factor of 1/180, for better display)

\vskip 0.5cm{\bf Figure 11} -- Typical contour plots of the joint probability
density function $p(Z_2,\Delta t_2; Z_1,\Delta t_1)$ of the monthly values signal
of the SOI from Jan. 1866 to Jan. 2006. Dashed lines have a slope +1 and
emphasize the correlations between probability density functions for $\Delta
t_2=1$~month and $\Delta t_1=2$~months. Contour levels correspond to
$log_{10}p(Z_2,\Delta t_2; Z_1,\Delta t_1) = -1.8,-2.0,-2.2,-2.4,-2.6,-2.8$ from
center to border

\vskip 0.5cm{\bf Figure 12} -- Kramers-Moyal drift (a) $D^{(1)}$ and diffusion
(b) $D^{(2)}$ coefficients as a function of normalized variability $Z$ (dots) and
non normalized variability $Z$ (open circles) of the monthly values signal of the
SOI; $ D^{(1)} = -0.37 Z - 0.01$ ((a) dots), $ D^{(2)} = 0.10 Z^2 - 0.10 Z +
0.33$ ((b) dots); $ D^{(1)} = -0.52 \Delta y - 0.02$ ((a) open circles), $
D^{(2)} = 0.20 \Delta y^2 - 0.10 \Delta y + 0.24$ ((b) open circles)

\newpage

\newpage \begin{table} \caption{Values of the scaling exponent from the DFA
analysis of normalized variability $Z$ for different values of the time lag
$\Delta t=1,3,6,12,24,36$~months, and crossover `box size' $n_x$ }
\begin{center} \begin{tabular}{rccc} \hline $\Delta t$& $1+H_{DFA1}$& $1+H_{DFA2}$&
$n_x$ \\ \hline 1 & 0.264$\pm$ 0.017 & 0.058$\pm$ 0.016
& 23\\ 3 & 0.594$\pm$ 0.038 & 0.082$\pm$ 0.060
& 23 \\ 6 & 0.909$\pm$ 0.028 & 0.101$\pm$ 0.063
& 23\\ 12 & 1.148$\pm$ 0.032 & 0.129$\pm$ 0.070 & 25 \\ 24 & 1.115$\pm$ 0.030 &
0.142$\pm$ 0.066 & 32 \\ 36 & 1.082$\pm$ 0.027 & 0.192$\pm$ 0.044 & 35 \\

\hline \end{tabular} \end{center} \label{table1} \end{table}

\begin{table} \caption{ Values of the parameters characterizing the monthly
values of the Southern Oscillation Index (SOI) from Jan. 1866 to Jan. 2006 in the
nonextensive thermostatistics approach. For the definition of the
Kolmogorov-Smirnov distance $d_{KS}$ see the text} \begin{center} \tabcolsep=4pt
\begin{tabular}{rcccccc} \hline $\Delta t$& $q$ &$\alpha$&$C\beta_0$& $p_{\Delta
t}(Z=0)$ & $K_r$&$d_{KS}$ \\ &&&&Eq.(4)&&\\ \hline 1 & 1.25 & 0.93 & 0.70 &
0.448 & 5 & 0.005 \\ 3 & 1.20 & 0.91 & 0.65 & 0.432 & 4.20 & 0.012 \\ 6
& 1.16 & 0.90 & 0.63 & 0.426 & 3.80 & 0.009 \\ 12 & 1.10 & 0.88 & 0.60 &
0.415 & 3.40 & 0.008 \\ 24 & 1.06 & 0.87 & 0.58 & 0.407 & 3.21 & 0.009 \\ 36
& 1.01 & 0.87 & 0.56 & 0.402 & 3.03 & 0.010 \\ \hline \end{tabular}
\end{center} \label{table2} \end{table}

\newpage

\begin{figure}[ht] \begin{center} \leavevmode \epsfysize=10cm
\epsffile{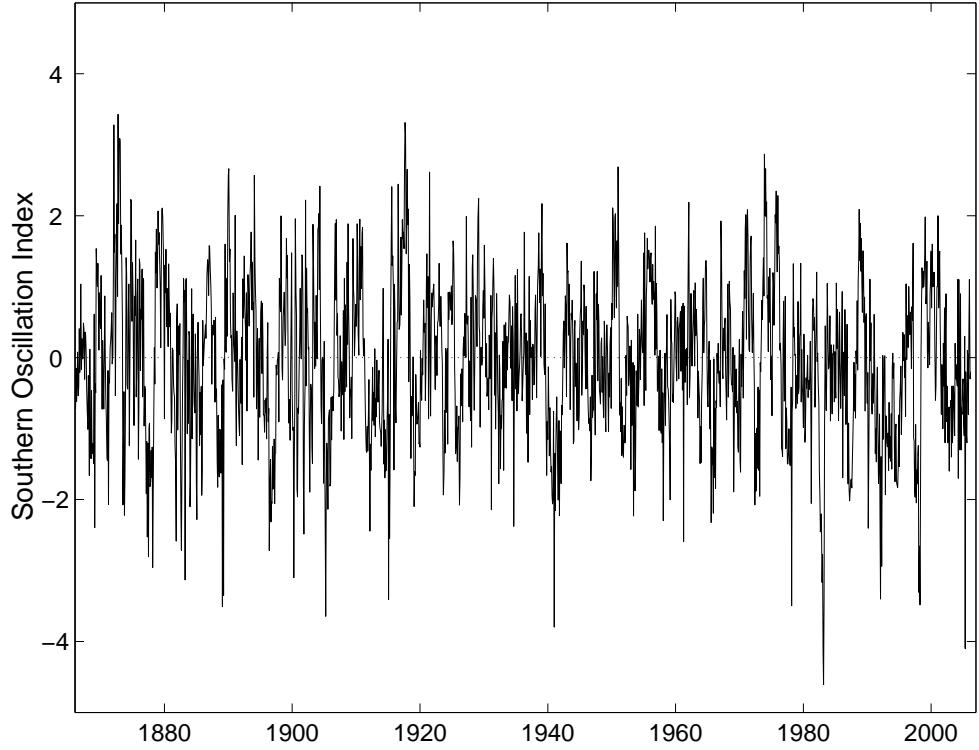} \caption{Monthly values of the Southern Oscillation Index
(SOI) as defined in the text reported from Jan. 1866 to Jan. 2006. Data are
downloaded from
$http://tao.atmos.washington.edu/pacs/additional_{}analyses/soi.html$ and from
$ftp://ftpprd.ncep.noaa.gov/pub/cpc/wd52dg/data/indices/soi$ after June 1999.
Data series consists of 1681 data points} \end{center} \label{fig1}\end{figure}

\begin{figure}[ht] \begin{center} \leavevmode \epsfysize=10cm
\epsffile{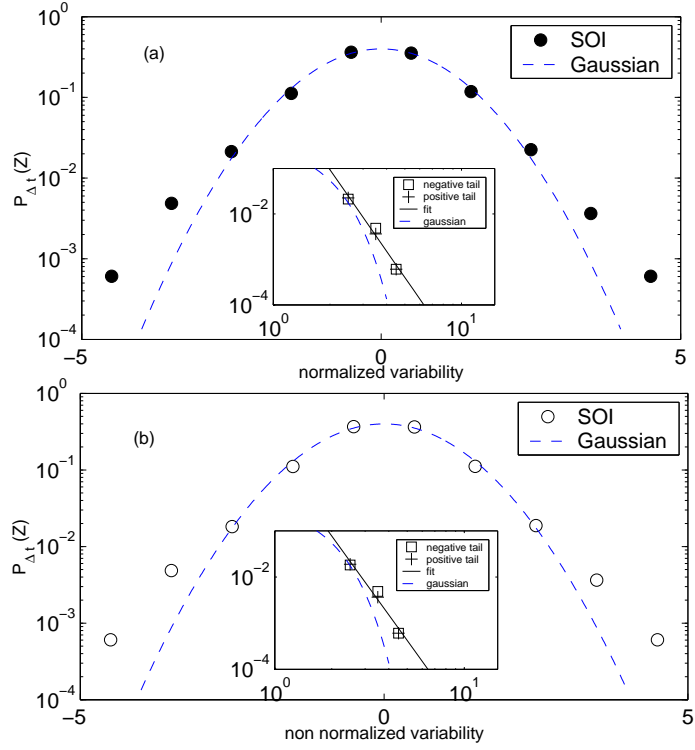}
\caption{(a) Probability distribution function of normalized variability
$Z(t,\Delta t)$ of monthly values signal of the Southern Oscillation Index from
Jan. 1866 to Jan. 2006 for $\Delta t=1$~month (symbols). $Z(t,\Delta t)$ is
defined as $Z(t,\Delta t)=(\Delta y(t) - <\Delta y>_{\Delta t})/\sigma_{\Delta
t}$, where $\Delta y(t)=y(t+\Delta t)-y(t)$ and $\sigma_{\Delta t}$ is the
standard deviation of $\Delta y(t)$ for time lag $\Delta t$. The dashed line
represents a Gaussian distribution. Inset: Power law fit (solid line) of the
negative and positive slope (-5.6 for both) of the distribution outside the
Gaussian regime, i.e. $]-2,+2[$. (b) same as (a) but for non normalized
variability $\Delta y$} \end{center} \label{fig1}\end{figure}

\newpage \begin{figure}[ht] \begin{center} \leavevmode \epsfysize=10cm
\epsffile{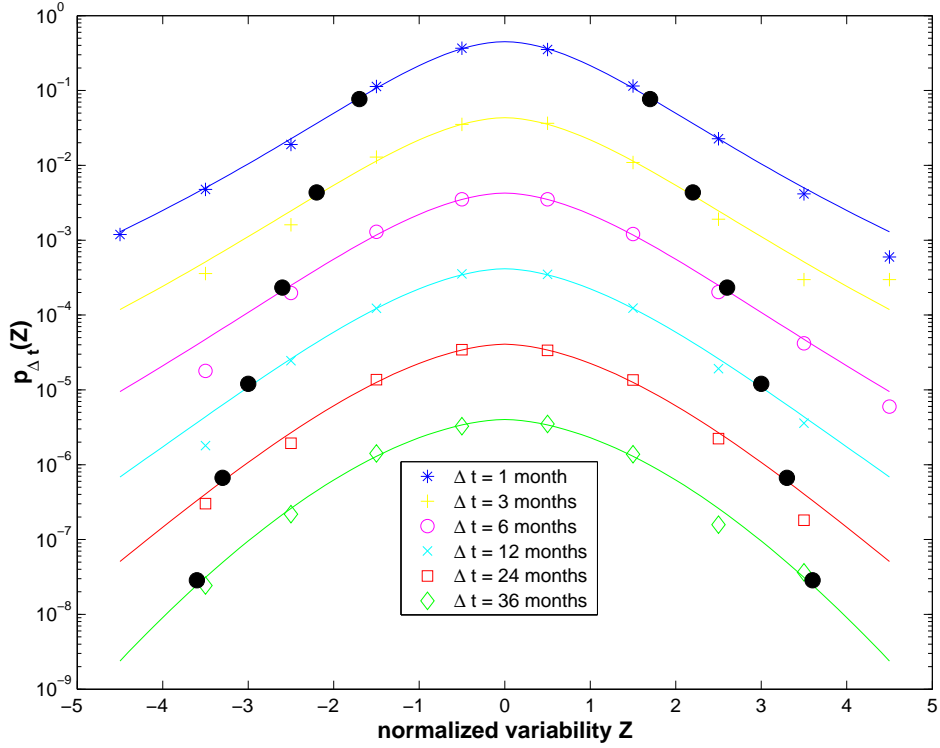} \caption{Probability distribution function (PDF) $p_{\Delta
t}(Z)$ of normalized variability of monthly values signal of the SOI (symbols)
and the Tsallis type distribution function (lines) for different values of
$\Delta t=1,3,6,12,24,36$~months. The PDF (symbols and curves) for each $\Delta
t$ are moved down by 10 with respect to the previous one; the curve for $\Delta
t=1$~month is unmoved. The large dots mark the ends of the interval in which the
distribution is like a gaussian distribution. The values of the parameters for
the Tsallis type distribution function for each $\Delta t$ are summarized in
Table \ref{table2}} \end{center}\label{fig2} \end{figure}

\newpage \begin{figure}[ht] \begin{center} \leavevmode \epsfysize=10cm
\epsffile{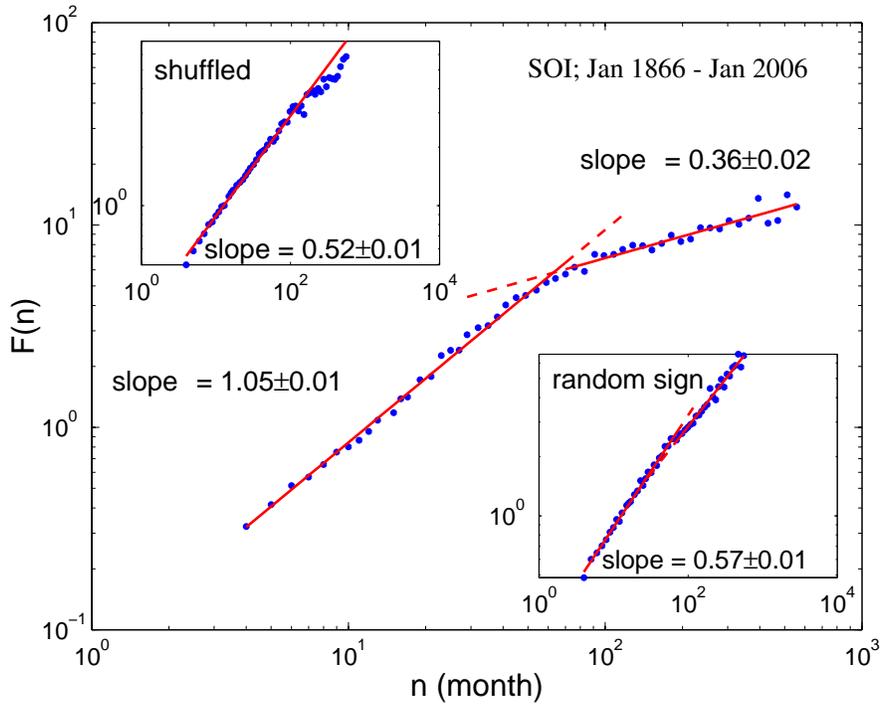} \caption{DFA function $F(n)$ plotted as a function of the
the box size $n$ of the integrated monthly values signal of the SOI from Jan.
1866 to Jan. 2006. $1/f$-like fluctuations with $slope=1.06\pm0.01$ are
obtained for time scales below 66 months and fractional Gaussian noise like
fluctuations $slope=0.36\pm0.02$ above 72 months. Insets: White noise like
fluctuations of two types of surrogate data, when the data are shuffled randomly
and when the sign of the data is shuffled randomly} \end{center}
\label{fig3}\end{figure}

\newpage \begin{figure}[ht] \begin{center} \leavevmode \epsfysize=10cm
\epsffile{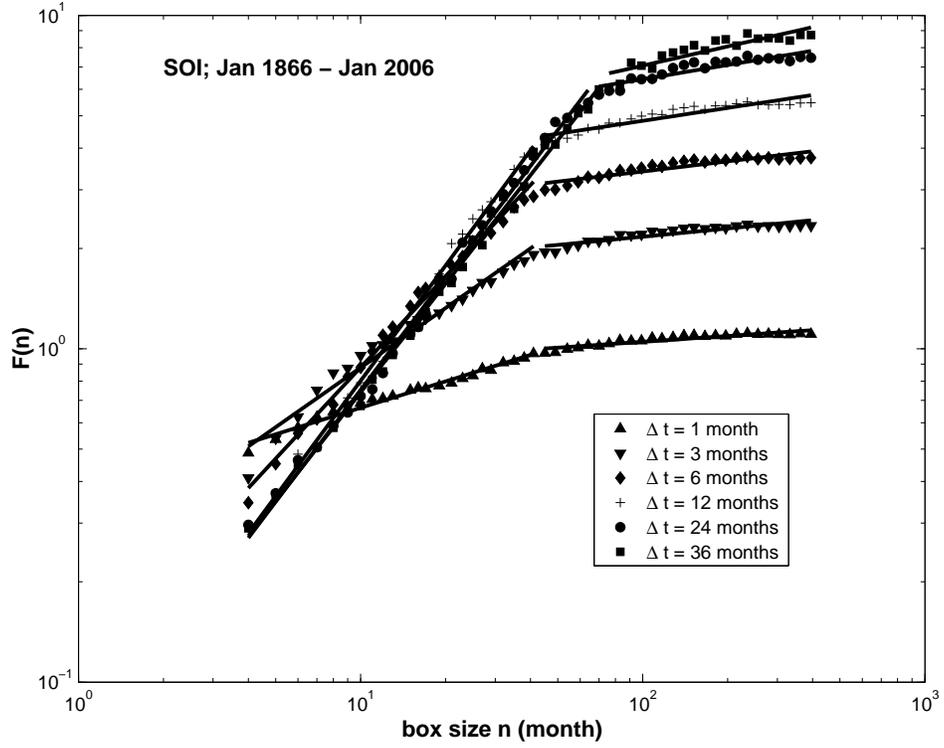} \caption{DFA function $F(n)$ plotted as a function of the
box size $n$ of the integrated normalized variability $Z(t,\Delta t)$ of the
monthly values signal of the SOI from Jan. 1866 to Jan. 2006, for different time
lags $\Delta t=1,3,6,12,24,36$~months. Values of the scaling exponents $H_{DFA}$
for the various DFA functions are summarized in Table \ref{table1}} \end{center}
\label{fig4}\end{figure}

\newpage \begin{figure}[ht] \begin{center} \leavevmode \epsfysize=10cm
\epsffile{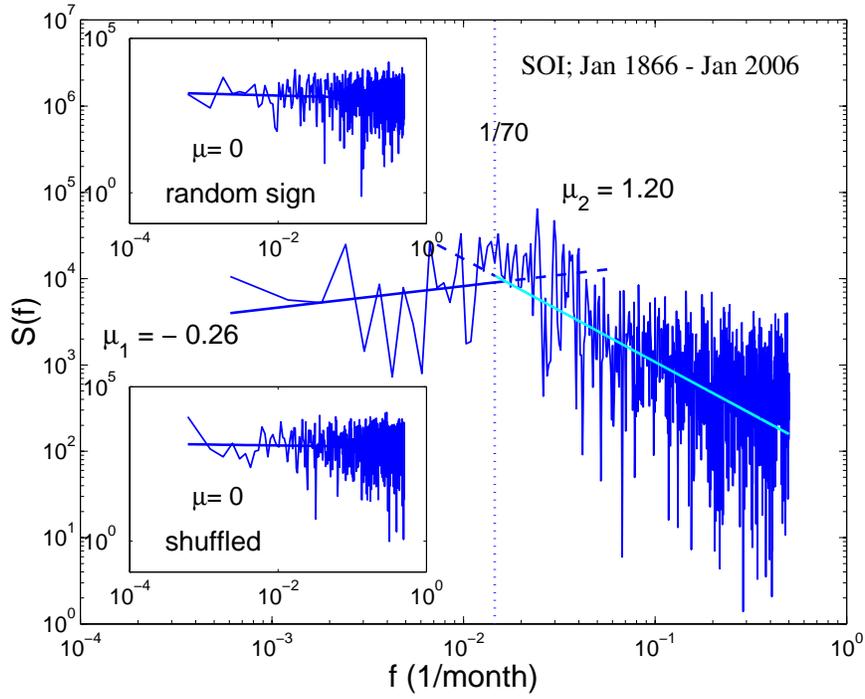} \caption{Power spectrum $S(f)$ of the monthly values signal
of the SOI from Jan. 1866 to Jan. 2006. A scale break at around
$f=1/70$~month$^{-1}$ separates two scaling regions. Insets: Scaling of the power
spectrum of both shuffled amplitude and shuffled sign of monthly values signal of the SOI
as a white noise signal with $\mu\approx 0$} \end{center}
\label{fig5}\end{figure}

\newpage \begin{figure}[ht] \begin{center} \leavevmode \epsfysize=10cm
\epsffile{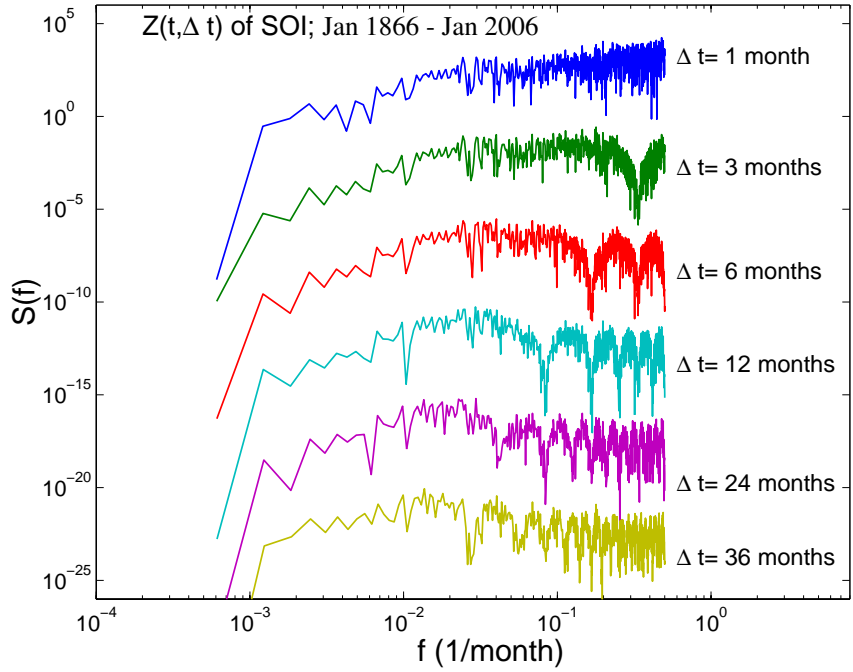} \caption{Power spectrum $S(f)$ of the normalized variability
$Z(t,\Delta t)$ of the monthly values signal of the SOI from Jan. 1866 to Apr.
2003 for different time lags $\Delta t=1,3,6,12,24,36$~months. Each curve is
moved down by $10^{-5}$ with respect to the previous one; the power spectrum of
the normalized returns for $\Delta t=1$~month is not displaced}
\end{center}\label{fig6} \end{figure}

\begin{figure}[ht] \begin{center} \leavevmode \epsfysize=8cm
\epsffile{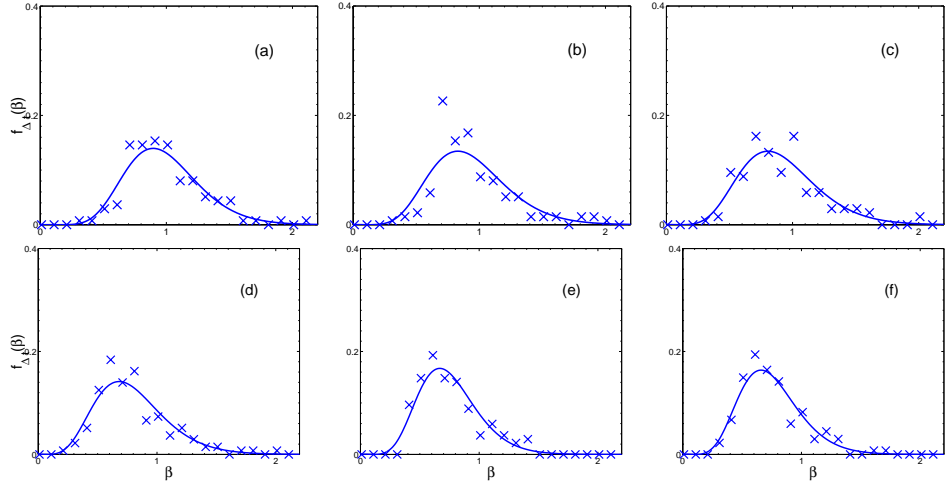}  \end{center} \caption{Probability density
$f_{\Delta t}(\beta)$ of the local volatility $\beta$ (Eq.(\ref{volatility})) in
terms of standard deviation of the normalized variability $Z(t,\Delta t)$ of SOI
in non-overlapping windows with size $m$=12~months for different time lags
(symbols) (a-f) $\Delta t=1,3,6,12,24,36$~months. Lines: $\chi^2$-distribution as
given by Eq. (\ref{chi})} \label{fig7}\end{figure}

\newpage \begin{figure}[ht] \begin{center} \leavevmode \epsfysize=10cm
\epsffile{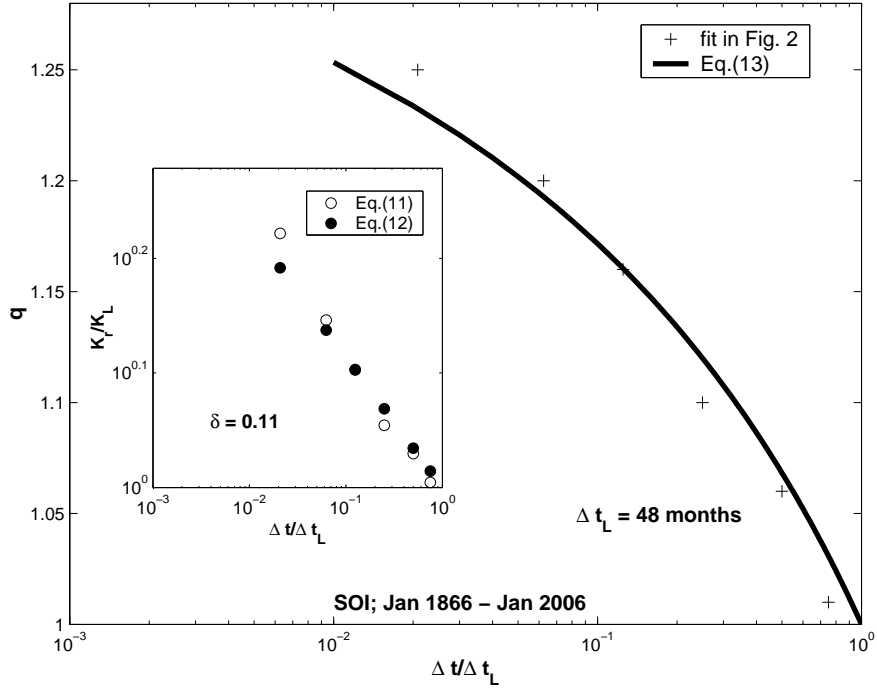} \caption{The functional dependence of the Tsallis $q$
parameter on the rescaled time lag $\Delta t/\Delta t_L$ for $\Delta
t_L=48$~months and $\delta= 0.11$ (see Eq. (\ref{qkr})) (line); the symbols
represent the values of the $q$ parameter listed in Table \ref{table2} and used
to plot the fitting lines in Fig. 2. Inset : Scaling properties of the rescaled
kurtosis $K_r/K_L$, where $K_L=3$ is the kurtosis for a Gaussian process, as a
function of the rescaled time lag $\Delta t/\Delta t_L$ satisfying Eq. (\ref{kr})
(open symbols) and Eq. (\ref{krL}) (full symbols)} \end{center}\label{fig8}
\end{figure}

\newpage \begin{figure}[ht] \begin{center} \leavevmode \epsfysize=10cm
\epsffile{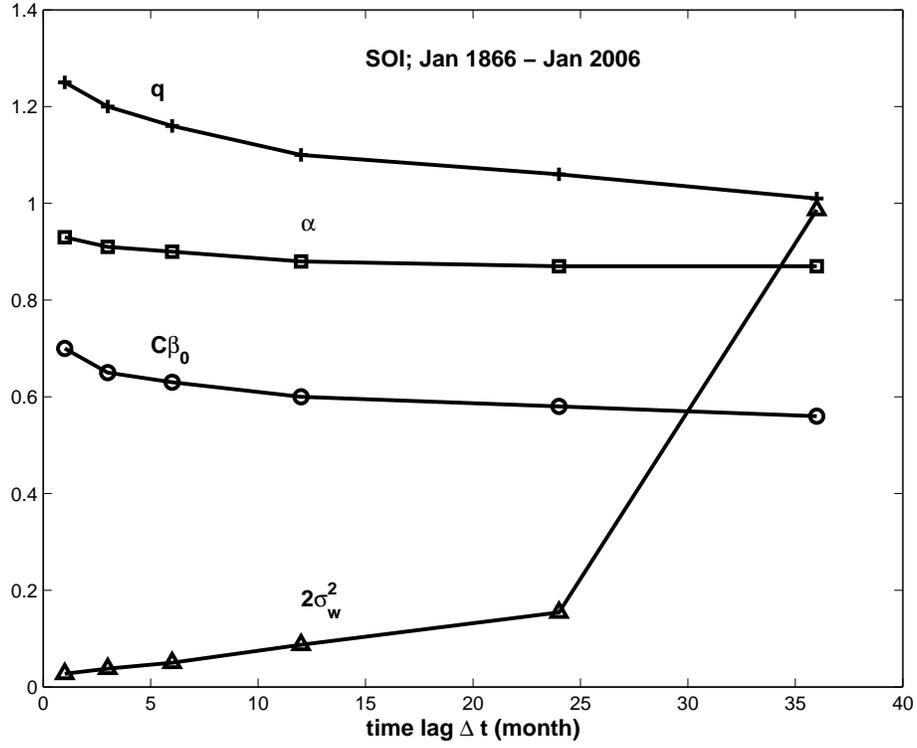} \caption{Characteristic parameters of Tsallis type
distribution function as defined in \cite{kozuki} : Tsallis $q$-parameter
(crosses), $\alpha$ (squares), constant $C\beta_0$ used in the fit (open
circles), the width of the Tsallis type distribution $2\sigma_w^2=(2\alpha-(q-1))
/ (2\alpha C\beta_0(q-1))$ from Eq.(\ref{tsallis}) (triangles) (rescaled by a
factor of 1/180, for better display) } \end{center}\label{fig9} \end{figure}

\begin{figure}[ht] \begin{center} \leavevmode \epsfysize=10cm
\epsffile{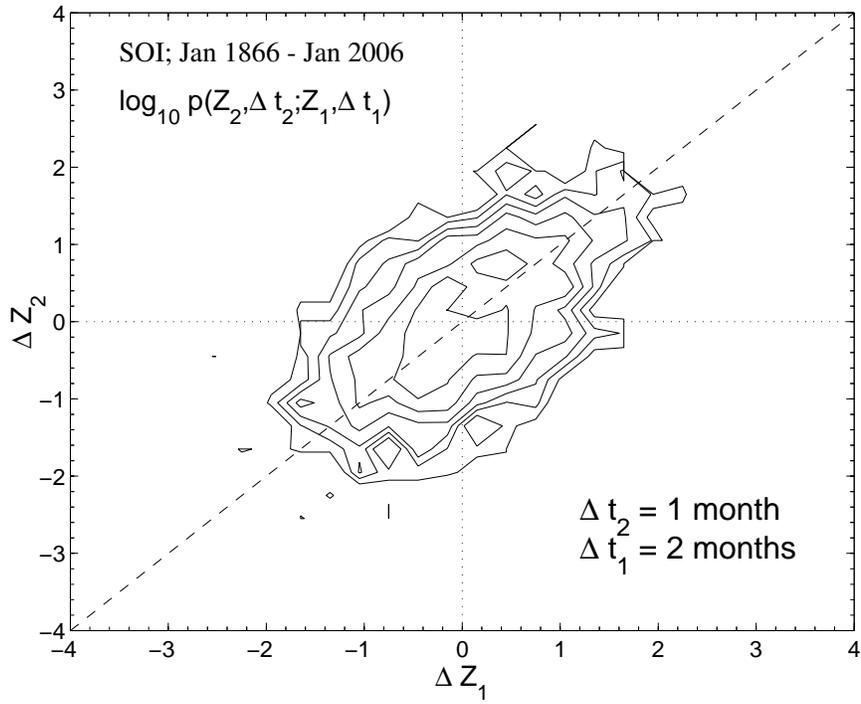} \end{center} \caption{Typical contour plots of the joint
probability density function $p(Z_2,\Delta t_2; Z_1,\Delta t_1)$ of the monthly
values signal of the SOI from Jan. 1866 to Jan. 2006. Dashed lines have a slope
+1 and emphasize the correlations between probability density functions for
$\Delta t_2=1$~month and $\Delta t_1=2$~months. Contour levels correspond to
$log_{10}p(Z_2,\Delta t_2; Z_1,\Delta t_1) = -1.8,-2.0,-2.2,-2.4,-2.6,-2.8$ from
center to border}\label{fig10} \end{figure}

\begin{figure}[ht] \begin{center} \leavevmode \epsfysize=12cm \epsfxsize=10cm
\epsffile{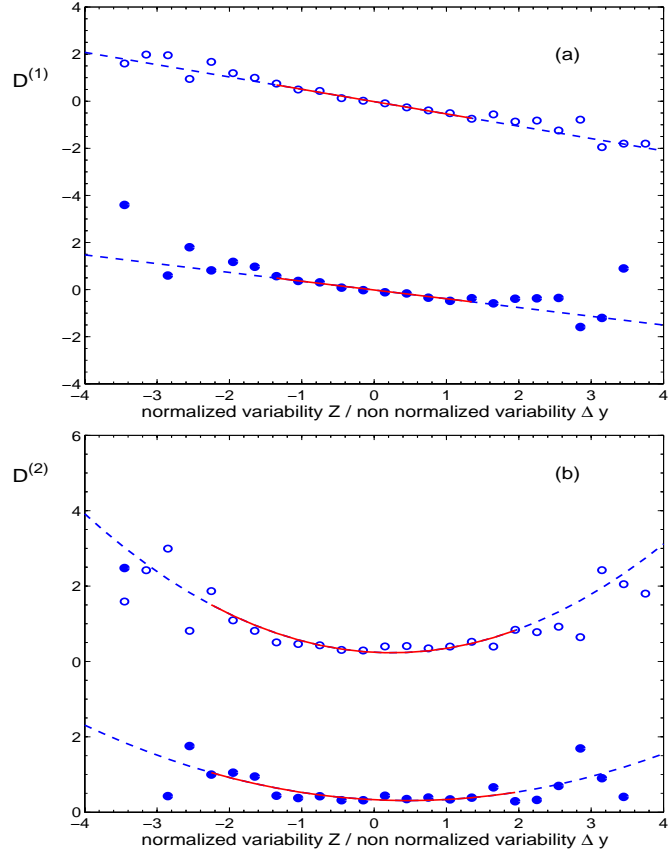}
\end{center} \caption{Kramers-Moyal drift (a) $D^{(1)}$ and diffusion (b)
$D^{(2)}$ coefficients as a function of normalized variability $Z$ (dots) and non
normalized variability $Z$ (open circles) of the monthly values signal of the
SOI; $ D^{(1)} = -0.37 Z - 0.01$ ((a) dots), $ D^{(2)} = 0.10 Z^2 - 0.10 Z +
0.33$ ((b) dots); $ D^{(1)} = -0.52 \Delta y - 0.02$ ((a) open circles), $
D^{(2)} = 0.20 \Delta y^2 - 0.10 \Delta y + 0.24$ ((b) open
circles)}\label{fig11} \end{figure}

\end{document}